\documentclass[useAMS,usenatbib]{mn2e}
\usepackage{graphicx, color, setspace, ulem, natbib}
\title[Re--examining the Too--Big--To--Fail Problem]{Re--examining the Too--Big--To--Fail Problem for Dark Matter Haloes with Central Density Cores}
\author[Ogiya and Burkert]{Go Ogiya$^{1,2,3}$\thanks{E-mail:ogiya@mpe.mpg.de} and Andreas Burkert$^{1,2}$\thanks{Max--Planck Fellow}\\
$^{1}$Universit\"ats-Sternwarte M\"unchen, Scheinerstra\ss e 1, D-81679 M\"unchen, Germany\\
$^{2}$Max-Planck-Institut f\"ur extraterrestrische Physik, Postfach 1312, Giessenbachstra\ss e, D-85741 Garching, Germany \\
$^{3}$Graduate School of Pure and Applied Science, University of Tsukuba, 1-1-1, Tennodai, Tsukuba, Ibaraki, 305-8577, Japan
}
\begin{document}

\date{Accepted 2014 October 28. Received 2014 October 28; in original form 2014 August 27}

\pagerange{\pageref{firstpage}--\pageref{lastpage}} \pubyear{2014}

\maketitle

\label{firstpage}

\begin{abstract}
Recent studies found the 
densities 
of dark matter (DM) subhaloes which surround nearby dwarf spheroidal galaxies (dSphs) to be significantly lower than those of  the most massive subhaloes expected around Milky Way sized galaxies in cosmological simulations, the so called ``too--big--to--fail'' (TBTF) problem. 
A caveat of previous work has been that dark substructures were assumed to contain steep density cusps in the center of DM haloes even though the central density structure of DM haloes is still under debate. 
In this study, we re--examine the TBTF problem for models of DM density structure with cores or shallowed cusps. 
Our analysis demonstrates that the TBTF problem is alleviated as the logarithmic slope of the central cusp becomes shallower. 
We find that the TBTF problem is avoided if the central cusps of DM haloes surrounding dSphs are shallower than $r^{-0.6}$. 
\end{abstract}


\section{Introduction}
\label{sec:int}
The local universe is a good site to test cosmological models. 
The current standard paradigm, the $\Lambda$ cold dark matter ($\Lambda$CDM) model, reproduces the large--scale properties of the universe successfully \citep[e.g.,][]{2004ApJ...606..702T, 2005ApJ...633..560E, 2007ApJS..170..288H}. 
However, some serious discrepancies between the $\Lambda$CDM prediction and observations have been identified in the local universe and are known as still remaining small--scale problems of $\Lambda$CDM cosmology. 

For example, recent studies revealed that the 
densities 
of the most massive dark matter (DM) subhaloes expected around Milky Way (MW) sized haloes in cosmological dissipationless simulations are much 
higher 
than those of subhaloes which surround nearby 
classical dwarf spheroidal galaxies (dSphs), the so called too--big--to--fail (TBTF) problem \citep{2011MNRAS.415L..40B, 2012MNRAS.422.1203B, 2014arXiv1404.5313G}. 
Massive satellite haloes are more likely to host galaxies than less--massive haloes because of their deeper potential well. 
These theoretically expected massive satellites that are however not observed are called ``massive failures''. 

Three satellite galaxies more massive than dSphs exist around the MW, i.e. the Large-- and Small Magellanic Clouds \citep[e.g.][]{2006lgal.symp...47V, 2009MNRAS.395..342B} and the Sagittarius dwarf galaxy. 
The Sagittarius dwarf galaxy is interacting with the MW and the stripped stars are observed as the Sagittarius stellar stream \citep{2003ApJ...599.1082M}. 
\cite{2010ApJ...712..516N} estimate that the dynamical mass of the progenitor of the Sagittarius dwarf galaxy is $\sim 10^{10} M_{\rm \odot}$. 
Therefore, we define the condition to solve the TBTF problem as $N_{\rm mf} \leq 3$ in this study. 
The TBTF problem is argued not only for the MW but also for other galaxies \citep{2014arXiv1407.4665P}. 
Disturbances in gas disks may be a powerful tool to estimate the dynamical properties of satellites \citep{2011ApJ...743...35C}. 
This problem also means a crisis for the abundance matching technique which assumes that the stellar mass of galaxies is a monotonic function of halo mass. 

Independent of the TBTF problem, the 
controversy 
about the inner mass--density structure of DM haloes, the core--cusp problem, has been under debate for two decades. 
Cosmological $N$--body simulations, based on the CDM model, predict the existence of a divergent density 
distribution, a cusp, in the centre of haloes 
\citep[][and references therein]{1991ApJ...378..496D, 1997ApJ...490..493N, 1997ApJ...477L...9F, 1999MNRAS.310.1147M, 2000ApJ...529L..69J, 2010MNRAS.402...21N, 2013ApJ...767..146I}. 
On the other hand, galaxies dynamically dominated by DM such as dwarf and low--surface--brightness galaxies 
seem to 
disagree with such a cuspy mass--density structure and have a constant density core 
\citep[e.g.,][]{1994Natur.370..629M, 2001ApJ...552L..23D, 2003ApJ...583..732S, 2005AJ....129.2119S, 2009A&A...505....1V, 2011AJ....141..193O, 2014MNRAS.443.3712H}. 

The cuspy profiles are derived from DM only simulations and the fluctuations in the gravitational potential
due to baryonic dynamical processes are supposed to alter the inner structure of DM haloes. Such fluctuations
could be induced by violent gas outflows, driven by supernova- and/or AGN feedback 
\citep[][and references therein]{1996MNRAS.283L..72N, 2011ApJ...736L...2O, 2014ApJ...793...46O, 2012MNRAS.421.3464P, 2013MNRAS.429.3068T, 2014ApJ...782L..39A} 
or by dynamical friction of gas or stellar 
clumps, spiraling into the center \citep[e.g.,][]{2001ApJ...560..636E, 2004PhRvL..93b1301M, 2006ApJ...649..591T, 2010ApJ...725.1707G, 2011MNRAS.418.2527I}. 
\cite{2014MNRAS.440L..71O} showed that the cusp--to--core transformation reproduces observed scaling relations of DM cores well. 

The inner density structure of the TBTF haloes is still a matter to debate.
Some studies adovocate the existence of a density core in the center of some dSphs \citep[e.g.,][]{2011ApJ...742...20W, 2012ApJ...754L..39A, 2012ApJ...746...89J, 2012ApJ...755..145H, 2013MNRAS.429L..89A}. 
On the other hand, it is still difficult to distinguish cored profiles from cuspy ones 
mainly due to observational uncertainties in the central regions 
\citep[e.g.,][]{2007AJ....134..566K, 2009ApJ...704.1274W, 2013A&A...558A..35B, 2013arXiv1305.0670R, 2014arXiv1406.6079S}. 
These difficulties are expected to be solved by ongoing and forthcoming observations such as GAIA \citep{2012Ap&SS.341...31D} and the Subaru Hyper--Suprime--Camera \citep{2010AIPC.1279..120T}. 

Even though the inner density profile of dSphs is uncertain, previous work has assumed cuspy models for DM haloes. 
The question then arises how a more cored profile would affect the TBTF puzzle.
The TBTF problem is defined in the $V_{\rm max}-R_{\rm max}$ plane, where $V_{\rm max}$ is the maximum circular velocity defined as 
\begin{equation}
V_{\rm max} = \max{\bigr[ V_{\rm c}(r) \bigl]} = \max{\biggl[ \sqrt{\frac{G M(r)}{r}} \biggr]},  \label{vc}
\end{equation}
with $R_{\rm max}$ the radius at which $V_{\rm max}$ is attained, $G$ is the gravitational constant and $M(r)$ 
is the mass within radius $r$, respectively. 
A transformation of the DM mass profile should strongly affect the conclusion because $V_{\rm max}$ and $R_{\rm max}$ depend on the mass 
profile of DM halo models
\citep{2013MNRAS.431.1220D, 2013ApJ...777..119F, 2013MNRAS.433.3539G, 2013ApJ...765...38G, 2014arXiv1410.3825, 2014ApJ...789L..17M}. 

The motivation of this study is to re--examine the TBTF problem for models of DM density profiles with cores or shallowed cusps. 
We find that the TBTF problem is alleviated as the logarithmic slope of the central cusp becomes shallower. 
Our analysis demonstrates that for cored dark haloes the TBTF problem can be solved
and it provides the steepest, allowed logarithmic slope in order to avoid a TBTF problem.
This paper is organized as follows. The procedures and assumptions of the analysis are described in Section \ref{sec:ana}. In Section \ref{sec:res}, we present results of the analysis. 
We discuss and summarize the results in Section \ref{sec:dis} and \ref{sec:sum}, respectively. 

\section{analysis}
\label{sec:ana}
Following \citet[hereafter B11]{2011MNRAS.415L..40B}, we compare $V_{\rm max}-R_{\rm max}$ values constrained by observations with theoretical predictions. 

\subsection{Constraint by observations}
\label{subsec:obs}
For homogeneity with 
B11,  we constrain $V_{\rm max}$ and $R_{\rm max}$ by the kinetic data derived by \cite{2010MNRAS.406.1220W}, the deprojected half--light radii, $R_{\rm 1/2}$, and the dynamical masses within $R_{\rm 1/2}$ of dSphs, $M_{\rm 1/2}$. 
The data are consistent with the results of other studies \citep[e.g.][]{2011MNRAS.411.2118A}. 
DSphs are dynamically dominated by DM even within $R_{\rm 1/2}$ \citep[e.g.,][]{1998ARA&A..36..435M} and this property allows us to regard $M_{\rm 1/2}$ as DM mass. 
General models of DM mass--density profile are characterized by two parameters. For the 
Navarro--Frenk--White (NFW) model \citep{1997ApJ...490..493N}, 
\begin{equation}
\rho(r) = \frac{\rho_{\rm s} r_{\rm s}^3}{r (r + r_{\rm s})^2}, \label{nfw}
\end{equation}
these parameters are $\rho_{\rm s}$ and $r_{\rm s}$, the scale density and length, respectively. 
The mass profile is given by 
\begin{eqnarray}
M(r) = 4 \pi \rho_{\rm s} r_{\rm s}^3 \biggl [ \ln{ \biggl ( 1 + \frac{r}{ r_{\rm s}} \biggr )} - \frac{(r/r_{\rm s})}{1 + (r/r_{\rm s})} \biggr ], 
\label{nfw_mass}
\end{eqnarray}
for NFW haloes. 
Since only the enclosed mass within $R_{\rm 1/2}$ is obtained by observations, it is impossible to uniquely determine the characterictics for each DM halo. 

We assume not only NFW-type models but also more generalised DM density profiles to re--examine the TBTF problem. 
As described in Section \ref{sec:int}, the inner density structure of dSphs is an open question. 
One of the models which we apply to the analysis is the so called $\alpha$--model,  
\begin{equation}
\rho(r) = \frac{\rho_{\rm 0} r_{\rm 0}^3}{r^\alpha (r + r_{\rm 0})^{3-\alpha}}, \label{alpha}
\end{equation}
where $\alpha$, $\rho_{\rm 0}$ and $r_{\rm 0}$ mean the logarithmic slope of the central cusp and 
the scale density and length, respectively. 
Here, $\alpha = 1$ corresponds to the NFW model and the model of $\alpha = 0$ has a central core. 
In this study, we consider models which satisfy $0 \leq \alpha \leq 1$.  
The mass profile for the $\alpha$--model is 
\begin{eqnarray}
M(\alpha; r) &=& \frac{4 \pi \rho_0 r_0^3}{3 - \alpha} \biggl(\frac{r}{r_0} \biggr)^{3-\alpha} \nonumber \\ 
&\times& \,_2F_1 \biggl[ 3-\alpha, 3-\alpha, 4-\alpha; -\biggl(\frac{r}{r_0} \biggr) \biggr],  
\label{alpha_mass}
\end{eqnarray}
\citep{2013MNRAS.432.2837T}. 
Here, $_{\rm 2}F_{\rm 1} [3-\alpha, 3-\alpha, 4-\alpha; -(r/r_0)]$ is Gauss's hypergeometric function. 

We also check for the Burkert model 
\begin{equation}
\rho(r) = \frac{\rho_{\rm 0} r_{\rm 0}^3}{(r + r_{\rm 0}) (r^2 + r_0^2)}, \label{burkert}
\end{equation}
which is a cored model and well reproduces the mass--density structure of dwarf and spiral galaxies \citep{1995ApJ...447L..25B, 2000ApJ...537L...9S}. 
The mass profile is given by 
\begin{eqnarray}
M(r) = \pi \rho_0 r_0^3 \biggl [ -2 \arctan{\biggl ( \frac{r}{r_0} \biggr )} +2 \ln{\biggl \{ 1 + \biggl (\frac{r}{r_0} \biggr ) \biggr \}} \nonumber \\
+ \ln{\biggl \{ 1+ \biggl ( \frac{r}{r_0} \biggr )^2 \biggr \}} \biggr ], 
\label{burkert_mass}
\end{eqnarray}
for Burkert haloes \citep{2000ApJ...538..559M}. 

In the wake of B11, we consider nine satellite dwarf galaxies around the MW 
(Sextans, Canes Venatici I, Carina, Fornax, Leo I, Leo I\hspace{-.1em}I, Sculptor, Ursa Minor and Draco) 
to obtain the constraint from their kinetic data. 
1$\sigma$ confidence range is taken account for the observational data, $M_{\rm 1/2}$ and $R_{\rm 1/2}$. 
Using Eq. (\ref{vc}), we derive bands which represent acceptable values of $V_{\rm max}$ and $R_{\rm max}$ for respective satellites (see Fig. 1 of B11). 
The observational contraint is determined by combining the bands. 
In the plane of $V_{\rm max}$ and $R_{\rm max}$, the distribution of the observed dSphs of the MW is constrained within the shaded region.

To understand the results of the analysis, we need to consider the meaning of the circular velocity, $V_{\rm c}$ which
can be rewritten as 
\begin{equation}
V_{\rm c}(r) = \sqrt{G \frac{M(r)}{r}} \sim \sqrt{G \frac{dM(r)}{dr}} = \sqrt{4 G \pi r^2 \rho(r)}, \label{vc2}
\end{equation}
for spherical systems. For profiles with $\alpha > 0$, it increases around the center, reaches a peak at $R_{\rm max}$ 
where the density distribution is quasi-isothermal with a logarithmic slope of -2 and then declines in the ocutskirts where $\rho \propto r^{-3}$.  
$R_{\rm max}$ therefore is proportional to the scale length, $r_{\rm 0}$ and $V_{\rm max}$ is a function of the product, $\rho_{\rm 0} r_{\rm 0}^3$, 
since Eq. (\ref{vc2}) indicates the dependence, 
\begin{equation}
V_{\rm max} = V_{\rm c}(R_{\rm max}) \propto \sqrt{r^2 \rho_{\rm 0} r_{\rm 0}^3 r^{-2}} \propto \sqrt{\rho_{\rm 0} r_{\rm 0}^3}. 
\end{equation}

\subsection{Prediction for properties of dark haloes}
\label{subsec:the}
In order to compare the observational constraints with theoretical predictions
we assume that DM haloes form following an NFW profile initially. The structure of an NFW halo depends on the concentration
parameter $c \equiv r_{\rm 200} / r_{\rm s}$, where $r_{\rm 200}$ is the virial radius. 
Inside of $r_{\rm 200}$, the mean density of the DM halo is 200 times the critical density of the universe. 
The virial mass, $M_{200}$ is related to $r_{\rm 200}$ by $M_{200} \equiv (4 \pi / 3) 200 \rho_{\rm crit} (1+ z)^3 r_{\rm 200}^3$ where $\rho_{\rm crit}$ and $z$ are the critical density of the universe and redshift, respectively. 
The concentration parameter, $c$ is a function of $M_{\rm 200}$ and $z$.
We adopt $c(M_{\rm 200}, z)$ as proposed by \cite{2012MNRAS.423.3018P} which is appropriate down to $M_{\rm 200} \sim 10^8 M_{\rm \odot}$ which is the mass scale of dwarf galaxies \citep{2014MNRAS.440L..71O}. 

We then assume that the central cusp of the NFW halo is shallowed by some dynamical process and the density profile transforms into Eq. (\ref{alpha}) or (\ref{burkert}). We impose two physical conditions in order to determine the two free
parameters $\rho_{\rm 0}$ and $r_{\rm 0}$. 
The first one is the conservation of the virial mass, $M_{\rm 200} = M(r_{\rm 200})$. 
The second condition is the conservation of the mass--density in the outskirts. 
This is reasonable if the cusp shallowing is caused by fluctuations in the gravitational potential driven by 
baryonic flows in the inner dark halo regions. 
From equations (\ref{nfw}), (\ref{alpha}) and (\ref{burkert}), the following should then be satisfied, 
\begin{equation}
\rho_{\rm s} r_{\rm s}^3 = \rho_0 r_0^3. \label{outskirt}
\end{equation}
We also have to define a redshift at which the central cusp has been shallowed, $z_{\rm s}$, and assume that the parameters of DM haloes, $\rho_{\rm 0}$ and $r_{\rm 0}$, are conserved until a redshift, 
$z' < z_{\rm s}$.
In order to justify this assumption, we consider dark haloes which have not experienced significant mass growth in a time frame from $z_{\rm s}$ to $z' < z_{\rm s}$ (see below).

Fig. \ref{density_profiles} shows the resultant density profiles of DM haloes after the process of the cusp shallowing. 
The dark haloes have the identical initial NFW configuration. 
For the $\alpha$--model, the central cusp is shallower and the central density of DM haloes decreases the smaller $\alpha$. 
The core is even larger and the central density becomes even smaller
in Burkert haloes, compared to the $\alpha=0$ model. 

We now test the TBTF problem and its dependence on the DM halo profile by focussing on the satellite system of the MW.
\cite{1974ApJ...187..425P} have established a formalism to derive the number density of DM haloes for given halo mass and redshift. 
The predicted number densities well match the results of cosmological $N$--body simulations \citep[e.g.,][]{1999MNRAS.308..119S}. 
Subsequent studies extended the formalism and obtained useful expressions. 
\cite{1991MNRAS.248..332B} obtained a formula to compute the mass fraction of elements which were dark haloes of mass $M_{\rm 1}$ at $z_{\rm 1}$ and have merged to form a larger dark halo of mass $M_{\rm 0} > M_{\rm 1}$ by $z_{\rm 0} < z_{\rm 1}$. 
The average number of progenitors, $N_{\rm prg}$  is given by  
\begin{eqnarray}
N_{\rm prg} (M_{\rm 0}, z_{\rm 0} | M_{\rm 1}, z_{\rm 1}) &=& \sqrt{\frac{2}{\pi}} \frac{M_{\rm 0}}{M_{\rm 1}} \frac{\sigma_{\rm 1} (\delta_{\rm 1} - \delta_{\rm 0})}{(\sigma_{\rm 1}^2 - \sigma_{\rm 0}^2)^{3/2}}  \nonumber \\
                                                                                   &\times& \exp{\biggl [ -\frac{(\delta_{\rm 1} - \delta_{\rm 0})^2}{2(\sigma_{\rm 1}^2 - \sigma_{\rm 0}^2)} \biggr ]} \biggl | \frac{d \sigma_{\rm 1}}{dM_{\rm 1}} \biggr |,  
\label{n_prg}
\end{eqnarray}
where $\delta_{\rm n}$ and $\sigma_{\rm n}$ $({\rm n} = 0, 1)$ are the linear overdensity and the linear rms fluctuation of the density field, respectively. 
The linear overdensity, $\delta_{\rm n}$ is defined by $\delta_{\rm n} = \delta_{\rm c} / D(z_{\rm n})$ where $\delta_{\rm c} = 1.69$ is the critical overdensity to collapse and $D(z_{\rm n})$ is the linear growth factor measured at $z_{\rm n}$. 
The linear rms fluctuation of the density field, $\sigma_{\rm n}$ is a monotonically dicreasing function of halo mass. 

\cite{1993MNRAS.262..627L} derived a conditional probability, $P'(M', z' | M_{\rm 200}, z_{\rm s})$, that a dark halo makes a transition from $M = M_{\rm 200} < M'$ to $M > M'$ within a time frame from $z_{\rm s}$ to $z' < z_{\rm s}$.
The inverse probability which corresponds to the probability that a dark halo does not make the transition is given by $P(M', z' | M_{\rm 200}, z_{\rm s}) = 1 - P'(M', z' | M_{\rm 200}, z_{\rm s})$. 
$P$ can be regarded as the fraction of DM haloes with masses $M_{\rm 200}$ at $z_{\rm s}$ that survive until $z'$ without 
substantial growth of their mass 
and is defined by
\begin{eqnarray}
&P(M', z' | M_{\rm 200}, z_{\rm s}) = 1 - \frac{1}{2} \{1 - {\rm erf}(A) \}&  \nonumber\\
                                                &+ \frac{1}{2} \frac{\delta(z_{\rm s}) - 2 \delta(z')}{\delta(z_{\rm s})} \exp{\bigl [ \frac{2 \delta(z') \{ \delta(z_{\rm s}) - \delta(z') \}}{\sigma^2(z_{\rm s})}\bigr ]} \{1 -  {\rm erf}(B)\}, & 
\end{eqnarray}
where $A$ and $B$ are given by 
\begin{eqnarray}
&A& = \frac{\sigma^2(z_{\rm s}) \delta(z') - \sigma^2(z') \delta(z_{\rm s})}{\sqrt{2 \sigma^2(z_{\rm s}) \sigma^2(z') \{ \sigma^2(z_{\rm s}) - \sigma^2(z') \} }} \\
&B& = \frac{\sigma^2(z') \{ \delta(z_{\rm s}) - 2 \delta(z') \} + \sigma^2(z_{\rm s})\delta(z')}{\sqrt{2 \sigma^2(z_{\rm s}) \sigma^2(z') \{\sigma^2(z_{\rm s}) - \sigma^2(z') \} }}. 
\end{eqnarray}

We compute the number of mass elements contained in the MW halo which were DM haloes with $M_{\rm 200}$ at $z_{\rm s}$ by using Eq. (\ref{n_prg}), i.e. $N_{\rm prg} (M_{\rm MW}, 0 | M_{\rm 200}, z_{\rm s})$. 
The surviving satellite haloes in the MW halo are defined as dark haloes whose mass does not reach $M' = 2 M_{\rm 200}(z_{\rm s})$ until $z'=0$. 
We assume that DM haloes which satisfy this condition are surviving as independent haloes and halo properties have been conserved. 
The number of subhaloes of the MW with halo mass, $M_{\rm 200}$, and redshift for cusp shallowing, $z_{\rm s}$, is calculated by 
\begin{eqnarray}
N_{\rm 0}(M_{\rm 200}, z_{\rm s}) = N_{\rm prg} (M_{\rm MW}, 0 | M_{\rm 200}, z_{\rm s})  P(2M_{\rm 200}, 0|M_{\rm 200}, z_{\rm s}). \label{n0}
\end{eqnarray}

In order to compute $N_{\rm prg}$ and $P$, some cosmological parameters and the dynamical mass of the MW, $M_{\rm MW}$ are required. 
The cosmological parameters in the following analysis are determined by \cite{2011ApJS..192...18K}. 
For the MW, we adopt the dynamical mass $M_{\rm MW} = 2.43 \times 10^{12} M_{\rm \odot}$ \citep{2008MNRAS.384.1459L}. 

\begin{figure}
 \centering 
   \includegraphics[width=90mm]{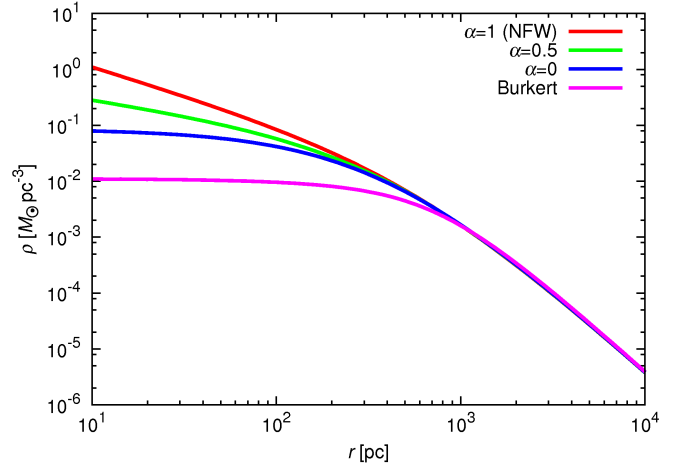}
     \caption{
         Density profile of DM haloes with $M_{\rm 200}=10^8 M_{\rm \odot}$ and $z_{\rm s}=0$ after the process of the cusp shallowing. 
       \label{density_profiles}
     }
\end{figure}

\section{Results}
\label{sec:res}

\begin{figure*}
 \centering 
   \includegraphics[width=120mm]{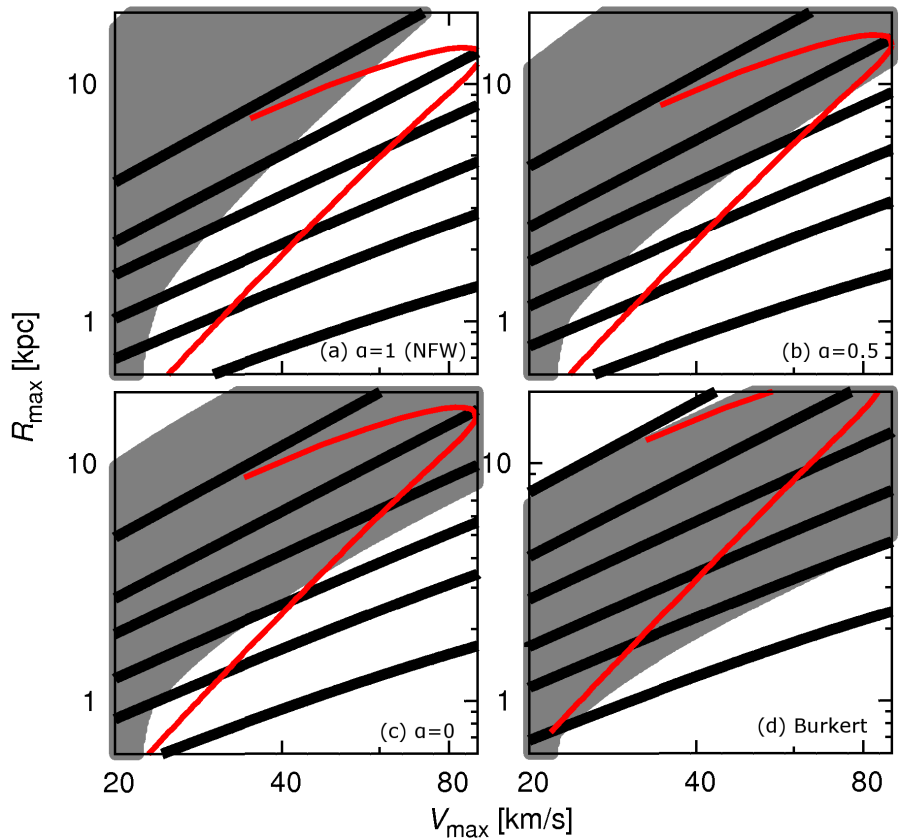}
     \caption{
         Comparison between observationally constrained and theoretically predicted properties of DM haloes in the $V_{\rm max}-R_{\rm max}$ plane. 
         Each panel depicts the results assuming various models of DM density profiles. 
         Shaded regions represent the locations of observed dSphs around the MW. 
         Black lines are the predicted properties of DM haloes assuming NFW haloes transformed 
         into mass--density models with central cores or shallowed cusps. 
           From the top black lines to the bottom ones, they show results for redshifts $z_{\rm s} = 0, 1, 3, 5, 7$ and $10$, respectively. 
           For given $V_{\rm max}$, $R_{\rm max}$ increases with decreasing $z_{\rm s}$.
          Red lines represent the contours where the expected number of dark haloes around the MW approaches unity: $N_{\rm 0} = 1$.
     \label{v_r_map}     
}
\end{figure*}

In order to re--examine the TBTF problem for DM density models with central cores or shallowed cusps, we parametrise the inner density structure of DM haloes in the analysis. 
In Fig. \ref{v_r_map}, we compare the constraints on DM subhaloes obtained from 
kinematic data of nearby dSphs with the predicted properties of DM haloes for various models of DM density profiles. 
Each panel shows the results for an NFW model with $\alpha=1$, $\alpha=0.5, 0$ and a Burkert profile. The observed
dSphs of the MW lie within the shaded regions. 
Red lines show the contour where of expected number of satellites around the MW at the present time is unity, i.e. $N_{\rm 0} = 1$ (see equation \ref{n0}). 
Black lines represent the theoretically predicted properties of dark haloes with inner profiles as introduced in Section \ref{sec:ana}. 
DM haloes with higher $z_{\rm s}$ are more compact than ones with lower $z_{\rm s}$ and located below lower $z_{\rm s}$ haloes in the $V_{\rm max}-R_{\rm max}$ plane. 
Considering at the given $z_{\rm s}$, dark haloes with smaller masses which are more abundant locate on the left side of ones with larger masses in the $V_{\rm max}-R_{\rm max}$ plane. 
Hence, satellites should be detected in the shaded regions enclosed by red lines and top black lines, results of $z_s = 0$, statistically. 
The size of the unshaded 
region 
enclosed by the red line and the lower bound of the shaded region 
correlates with the number of massive failures
, i.e. we find a large likelihood for massive failures for models with a steep central cusp. More quntitative discission is made with Fig. 3.

Panel (a) corresponds to NFW profiles and confirms the results of previous studies (cf. Fig. 2 of B11). 
The distribution of the observed dSphs is constrained within the shaded region. 
They however do not extend into the unshaded region above the red line. 
Fig. \ref{v_r_map} also demonstrates that 
the area 
enclosed by red line and the lower bound of the shaded region 
decreases, i.e., the TBTF problem is alleviated, as the central cusp is shallowed and the central density of DM haloes decreases. 

The black lines in Fig. \ref{v_r_map} show the expected correlation of $R_{\rm max}$ versus $V_{\rm max}$ for satellites
with various redshifts at which the central cusp has been shallowed, $z_{\rm s}$. 
As discussed 
in Section \ref{sec:ana}, 
the maximum circular velocity, $V_{\rm max}$ depends on the product, $\rho_{\rm 0} r_{\rm 0}^3$. 
This is one of the conditions imposed on the core formation process: the conservation of the mass--density in the outskirts of 
DM haloes, Eq. (\ref{outskirt}). 
Therefore, $V_{\rm max}$ is almost conserved during the shallowing process.
Shallowing the central cusps leads to an expansion of the central region of DM haloes. 
As a consequence, the radius $R_{\rm max}$ where the logarithmic slope of the density profile equals the isothermal value
of -2 moves outward. The amount of change in $R_{\rm max}$ increases with increasing
difference between the initial cuspy profile and the resultant core or shallowed cusp profile.

The shaded regions show the location of observed dSphs. 
They move to smaller $R_{\rm max}$  as the logarithmic slope of the central cusp becomes shallower. 
The DM haloes have the same $V_{\rm max}$, i.e., approximately the 
same $\rho_{\rm 0} r_{\rm 0}^3$ values for different density profiles. 
In order to satisfy the condition that the mass, enclosed within the half--light--radii $M_{\rm 1/2}$, is as observed,
models with central cores or shallow cusps need higher 
scale densities, $\rho_{\rm 0}$ 
than those with steep cusps. 
This leads the scale length of DM haloes, $r_{\rm 0}$, to decrease. 
Since $R_{\rm max}$ is proportional to $r_{\rm 0}$, the range of $R_{\rm max}$ occupied by observed dSphs decreases 
as the logarithmic slope of the central cusp becomes shallower. 

Next, we estimate the number of massive failures around the MW at the present time, $N_{\rm mf}$, and derive the critical logarithmic slope to solve the TBTF problem, $\alpha_{\rm crit}$. 
By integrating $N_{\rm 0}(M_{\rm 200}, z_{\rm s})$ in the unshaded region below the shaded ones, the number of massive failures can be calculated, 
\begin{eqnarray}
N_{\rm mf} = \int N_{\rm 0}(M_{\rm 200}, z_{\rm s}) d\ln{S},  
\end{eqnarray}
where $d \ln{S}$ is an element area in the logarithmic $V_{\rm max}-R_{\rm max}$ plane. 
This definition of massive failures follows B11 and \cite{2012MNRAS.422.1203B} and is referred to as ``strong massive failures'' in \cite{2014arXiv1404.5313G}.

\begin{figure}
 \centering 
   \includegraphics[width=90mm]{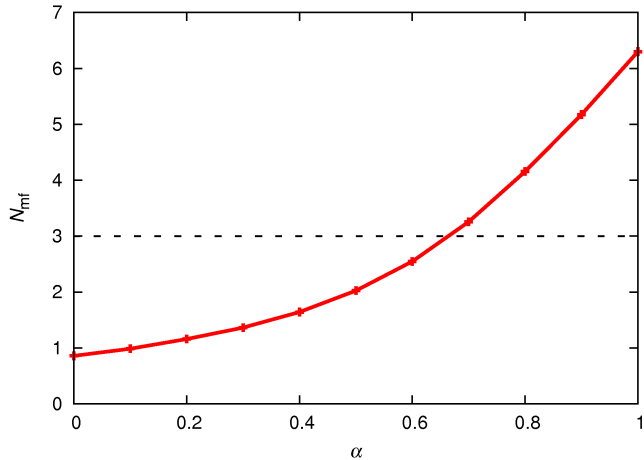}
     \caption{
         Number of massive failures,  $N_{\rm mf}$ as a function of the logarithmic slope of the central cusp, $\alpha$. 
         The black line corresponds to $N_{\rm mf}=3$. 
       \label{n_of_massive_failures}
     }
\end{figure}

Fig. \ref{n_of_massive_failures} demonstrates that the number of massive failures, $N_{\rm mf}$ decreases as the logarithmic slope of the central cusp, $\alpha$ becomes shallower. 
More than six massive failures exist around the MW if DM haloes follow NFW density profiles. 
This is consistent with the results of recent studies, based on numerical simulations \citep[cf. Fig. 3 of][]{2012MNRAS.422.1203B}. 
Fig. \ref{n_of_massive_failures} shows that $N_{\rm mf}$ falls below 3 in $\alpha < \alpha_{\rm crit} = 0.6$. 
If NFW haloes transform into Burkert haloes, $N_{\rm mf}$ decreases to 0.035 and the TBTF problem is completely solved. 

\section{Discussion}
\label{sec:dis}
\subsection{Potential of stellar feedback to solve the small--scale problems}
The results of the analysis indicate that the TBTF problem is closely connected to the
flattening of the central cusp, i.e., solving the core--cusp problem. 
Baryonic physics has been suggested as a unified solution for DM small--scale problems \citep{2014JCAP...04..021D}. 
A change in the gravitational potential around the center of galaxies driven by stellar feedback may be 
a promising
solution to decrease the central density of DM haloes. 

\cite{2012ApJ...759L..42P} calculated the required energy to transform cuspy haloes into cored haloes and 
compared it with the available energy from supernova feedback. 
Following this work, we define $\Delta W \equiv |W_{\rm ini} - W_{\rm fin}|/2$, where $W_{\rm ini}$ and $W_{\rm fin}$ is
the potential energy of the initial NFW halo and corresponding halo with central core or shallowed cusp, respectively. 
A Salpeter initial mass function \citep{1955ApJ...121..161S} is adopted to estimate the energy released by Type I\hspace{-.1em}I supernovae (SNe I\hspace{-.1em}I), $E_{\rm SN}$. The 
luminosity of dSphs around the MW is at least $10^5$ times the solar value \citep{1995MNRAS.277.1354I}. We therefore adopt
$10^5 M_{\rm \odot}$ for the stellar mass of galaxies and assume that a SN releases $10^{51} {\rm erg}$ of energy. 
The upper and lower mass limit of stars are set to $100 M_{\rm \odot}$ and $0.1 M_{\rm \odot}$, respectively. 
The lower mass limit of stars which will explode as SNe I\hspace{-.1em}I is assumed to be $8 M_{\rm \odot}$. 

\begin{figure}
 \centering 
   \includegraphics[width=90mm]{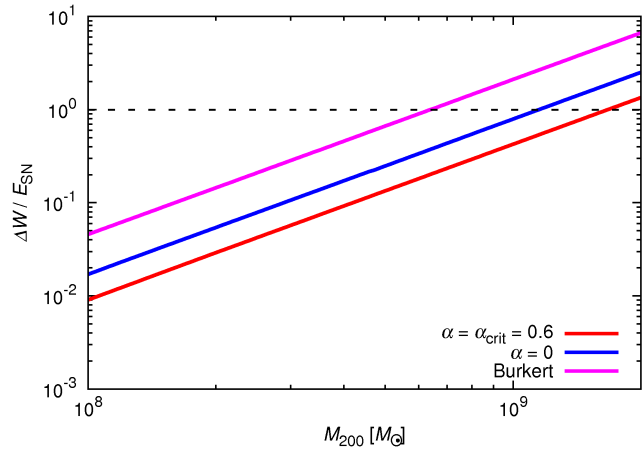}
     \caption{
  Ratio between the available energy of supernova feedback, $E_{\rm SN}$ and 
the required energy $\Delta W$ to transform cuspy haloes into haloes with central cores or shallowed cusps as a function of halo mass, $M_{\rm 200}$. 
        Red, blue and magenta lines represent results for models of $\alpha = \alpha_{\rm crit} = 0.6$, $\alpha = 0$ and Burkert profile, respectively. 
        The black, dashed line corresponds to $\Delta W = E_{\rm SN}$. 
         \label{energy_comparison}
     }
\end{figure}

According to 
some studies \citep[e.g.][]{2008MNRAS.386..348S, 2008Natur.454.1096S, 2011MNRAS.411.2118A}, 
the virial masses of nearby dSphs are $\sim 10^8 M_{\rm \odot}$. 
Here, we analyse DM haloes whose central cusps have been shallowed at $z_{\rm s} = 5$. 
Considering higher redshift, $\Delta W$ increases because DM haloes are denser than those of lower redshifts. 
The required energy, $\Delta W$ is approximately proportional to $(1 + z_{\rm s})$ due to the dependence of the virial radii of haloes on redshift. 
The dependence of $\Delta W$ on halo mass, $M_{\rm 200}$ is same with results of \cite{2014ApJ...782L..39A}, $\Delta W \propto M_{\rm 200}^{\rm 1.65}$.  
SN feedback provides haloes 
on and below 
the black, dashed line 
in Fig. \ref{energy_comparison} with sufficient energy to transform steep cusps into flat cores or shallowed cusps. 
This result highlights the potential of stellar feedback to solve the small--scale problems of $\Lambda$CDM cosmology. 
Whether this solution is reasonable depends critically on the fraction of SN energy that can be transferred to the 
DM haloes.
Fig. \ref{energy_comparison} indicates the required fraction of energy transferred into the DM distribution for each model. 
Despite of a lot of efforts this efficiency is still uncertain.  
More studies are needed in order to better understand the formation process of cores in dwarf
satellite haloes 
\citep[see also][]{2014ApJ...782L..39A}. 

\subsection{Dynamical mass of the Milky Way}
Another important constraint for the TBTF problem is the dynamical mass of the MW. 
Using the results of cosmological $N$--body simulations, \cite{2012MNRAS.424.2715W} demonstrate that the number of subhaloes which can be regarded as massive failures increases roughly linearly with the mass of the host haloes. 
Eq. (\ref{n_prg}) explains the dependence. 
\cite{2014arXiv1405.7697C} constrain the dynamical mass of the MW halo by the condition to match the number of observed satellite galaxies around the MW. 
The upper mass limit is determined from the condition to avoid the TBTF problem, i.e. the number of subhaloes which are more massive than dSphs should be equal to or less than 3. 

\begin{figure}
 \centering 
   \includegraphics[width=90mm]{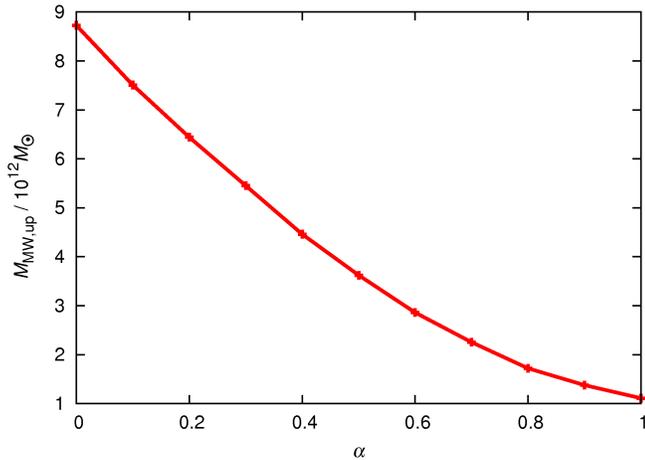}
     \caption{
         Upper mass limit of the MW halo derived from the condition to avoid the TBTF problem, $M_{\rm MW, up}$ as a function of the logarithmic slope of the central cusp, $\alpha$. 
       \label{mw_mass}
     }
\end{figure}

Following this condition, we derive the upper mass limit of the MW halo for respective density models of subhaloes. 
Fig. \ref{mw_mass} shows the upper mass limit of the MW halo required in order to avoid the TBTF problem, $M_{\rm MW, up}$ as a function of the logarithmic slope of the central cusp, $\alpha$. 
$M_{\rm MW, up}$ for the NFW model ($\alpha=1$), $\approx 1.1 \times 10^{12} M_{\rm \odot}$ is consistent with the constraint obtained by \cite{2014arXiv1405.7697C}. 
The allowed ranges of the MW halo mass are extended as the logarithmic slope of the central cusp becomes shallower. 
Therefore, a precise determination of the inner mass--density structure of satellite galaxies could provide interesting constraints for the host halo mass. 

\section{Summary}
\label{sec:sum}
In this study, we re--examined one of the small--scale problems in $\Lambda$CDM cosmology, the TBTF problem. 
The central density structure of DM haloes is still an open question for dSphs. 
Previous studies have assumed models of DM density profiles with steep cusps, such as the NFW profile. 
This motivated us to re--examine the problem for models of DM density structure with central cores or shallowed cusps. 
Our analysis demonstrates that the TBTF problem is alleviated as the logarithmic slope of the central cusp becomes shallower and 
it reduces to less than 3 failures for slopes shallower than $\alpha_{\rm crit} = 0.6$. 
Ongoing and forthcoming observational projects are expected to provide us data with sufficient quality to determine the 
values of $\alpha$ in the center of DM haloes, surrounding nearby dSphs. 
Combining the inner density structure of DM haloes with star formation histories of dwarf galaxies, we can get a better understanding of the core formation process \citep{2014ApJ...782L..39A, 2014MNRAS.437..415D, 2014arXiv1407.0022G, 2014ApJ...793...46O}.

\section*{Acknowledgments}
We are grateful to the anonymous referee for providing many helpful comments and suggestions. 
This work was supported in part by Grant--in--Aid for JSPS Fellows (25--1455 GO) and 
JSPS Grants--in--Aid for Scientific Research: (A) (21244013) and (C) (18540242). 
AB acknowledges support from the cluster of excellence ``Origin and Structure of the Universe''.

\bibliographystyle{mn2e}
\bibliography{./ref}

\begin{thebibliography}{80}
\expandafter\ifx\csname natexlab\endcsname\relax\def\natexlab#1{#1}\fi

\bibitem[{{Agnello} \& {Evans}(2012)}]{2012ApJ...754L..39A}
{Agnello} A., {Evans} N.~W., 2012, \apjl, 754, L39

\bibitem[{{Amorisco}, {Agnello} \& {Evans}(2013){Amorisco}, {Agnello}, \&
  {Evans}}]{2013MNRAS.429L..89A}
{Amorisco} N.~C., {Agnello} A., {Evans} N.~W., 2013, \mnras, 429, L89

\bibitem[{{Amorisco} \& {Evans}(2011)}]{2011MNRAS.411.2118A}
{Amorisco} N.~C., {Evans} N.~W., 2011, \mnras, 411, 2118

\bibitem[{{Amorisco}, {Zavala} \& {de Boer}(2014){Amorisco}, {Zavala}, \& {de
  Boer}}]{2014ApJ...782L..39A}
{Amorisco} N.~C., {Zavala} J., {de Boer} T.~J.~L., 2014, \apjl, 782, L39

\bibitem[{{Bekki} \& {Stanimirovi{\'c}}(2009)}]{2009MNRAS.395..342B}
{Bekki} K., {Stanimirovi{\'c}} S., 2009, \mnras, 395, 342

\bibitem[{{Bower}(1991)}]{1991MNRAS.248..332B}
{Bower} R.~G., 1991, \mnras, 248, 332

\bibitem[{{Boylan-Kolchin}, {Bullock} \& {Kaplinghat}(2011){Boylan-Kolchin},
  {Bullock}, \& {Kaplinghat}}]{2011MNRAS.415L..40B}
{Boylan-Kolchin} M., {Bullock} J.~S., {Kaplinghat} M., 2011, \mnras, 415, L40

\bibitem[{{Boylan-Kolchin}, {Bullock} \& {Kaplinghat}(2012){Boylan-Kolchin},
  {Bullock}, \& {Kaplinghat}}]{2012MNRAS.422.1203B}
{Boylan-Kolchin} M., {Bullock} J.~S., {Kaplinghat} M., 2012, \mnras, 422, 1203

\bibitem[{{Breddels} \& {Helmi}(2013)}]{2013A&A...558A..35B}
{Breddels} M.~A., {Helmi} A., 2013, \aap, 558, A35

\bibitem[{{Brook} \& {Di Cintio}(2014)}]{2014arXiv1410.3825}
{Brook} C., {Di Cintio} A., 2014, ArXiv e-prints

\bibitem[{{Burkert}(1995)}]{1995ApJ...447L..25B}
{Burkert} A., 1995, \apjl, 447, L25

\bibitem[{{Cautun} {et~al}\mbox{.}(2014){Cautun}, {Frenk}, {van de Weygaert},
  {Hellwing}, \& {Jones}}]{2014arXiv1405.7697C}
{Cautun} M., {Frenk} C.~S., {van de Weygaert} R., {Hellwing} W.~A., {Jones}
  B.~J.~T., 2014, ArXiv e-prints

\bibitem[{{Chakrabarti} {et~al}\mbox{.}(2011){Chakrabarti}, {Bigiel}, {Chang},
  \& {Blitz}}]{2011ApJ...743...35C}
{Chakrabarti} S., {Bigiel} F., {Chang} P., {Blitz} L., 2011, \apj, 743, 35

\bibitem[{{de Blok} {et~al}\mbox{.}(2001){de Blok}, {McGaugh}, {Bosma}, \&
  {Rubin}}]{2001ApJ...552L..23D}
{de Blok} W.~J.~G., {McGaugh} S.~S., {Bosma} A., {Rubin} V.~C., 2001, \apjl,
  552, L23

\bibitem[{{de Bruijne}(2012)}]{2012Ap&SS.341...31D}
{de Bruijne} J.~H.~J., 2012, \apss, 341, 31

\bibitem[{{Del Popolo} {et~al}\mbox{.}(2014){Del Popolo}, {Lima}, {Fabris}, \&
  {Rodrigues}}]{2014JCAP...04..021D}
{Del Popolo} A., {Lima} J.~A.~S., {Fabris} J.~C., {Rodrigues} D.~C., 2014,
  \jcap, 4, 21

\bibitem[{{Di Cintio} {et~al}\mbox{.}(2014){Di Cintio}, {Brook}, {Macci{\`o}},
  {Stinson}, {Knebe}, {Dutton}, \& {Wadsley}}]{2014MNRAS.437..415D}
{Di Cintio} A., {Brook} C.~B., {Macci{\`o}} A.~V., {Stinson} G.~S., {Knebe} A.,
  {Dutton} A.~A., {Wadsley} J., 2014, \mnras, 437, 415

\bibitem[{{Di Cintio} {et~al}\mbox{.}(2013){Di Cintio}, {Knebe}, {Libeskind},
  {Brook}, {Yepes}, {Gottl{\"o}ber}, \& {Hoffman}}]{2013MNRAS.431.1220D}
{Di Cintio} A., {Knebe} A., {Libeskind} N.~I., {Brook} C., {Yepes} G.,
  {Gottl{\"o}ber} S., {Hoffman} Y., 2013, \mnras, 431, 1220

\bibitem[{{Dubinski} \& {Carlberg}(1991)}]{1991ApJ...378..496D}
{Dubinski} J., {Carlberg} R.~G., 1991, \apj, 378, 496

\bibitem[{{Eisenstein} {et~al}\mbox{.}(2005){Eisenstein}, {Zehavi}, {Hogg},
  {Scoccimarro}, {Blanton}, {Nichol}, {Scranton}, {Seo}, {Tegmark}, {Zheng},
  {Anderson}, {Annis}, {Bahcall}, {Brinkmann}, {Burles}, {Castander},
  {Connolly}, {Csabai}, {Doi}, {Fukugita}, {Frieman}, {Glazebrook}, {Gunn},
  {Hendry}, {Hennessy}, {Ivezi{\'c}}, {Kent}, {Knapp}, {Lin}, {Loh}, {Lupton},
  {Margon}, {McKay}, {Meiksin}, {Munn}, {Pope}, {Richmond}, {Schlegel},
  {Schneider}, {Shimasaku}, {Stoughton}, {Strauss}, {SubbaRao}, {Szalay},
  {Szapudi}, {Tucker}, {Yanny}, \& {York}}]{2005ApJ...633..560E}
{Eisenstein} D.~J. {et~al.}, 2005, \apj, 633, 560

\bibitem[{{El-Zant}, {Shlosman} \& {Hoffman}(2001){El-Zant}, {Shlosman}, \&
  {Hoffman}}]{2001ApJ...560..636E}
{El-Zant} A., {Shlosman} I., {Hoffman} Y., 2001, \apj, 560, 636

\bibitem[{{Faerman}, {Sternberg} \& {McKee}(2013){Faerman}, {Sternberg}, \&
  {McKee}}]{2013ApJ...777..119F}
{Faerman} Y., {Sternberg} A., {McKee} C.~F., 2013, \apj, 777, 119

\bibitem[{{Fukushige} \& {Makino}(1997)}]{1997ApJ...477L...9F}
{Fukushige} T., {Makino} J., 1997, \apjl, 477, L9

\bibitem[{{Garrison-Kimmel} {et~al}\mbox{.}(2014){Garrison-Kimmel},
  {Boylan-Kolchin}, {Bullock}, \& {Kirby}}]{2014arXiv1404.5313G}
{Garrison-Kimmel} S., {Boylan-Kolchin} M., {Bullock} J.~S., {Kirby} E.~N.,
  2014, ArXiv e-prints

\bibitem[{{Garrison-Kimmel} {et~al}\mbox{.}(2013){Garrison-Kimmel}, {Rocha},
  {Boylan-Kolchin}, {Bullock}, \& {Lally}}]{2013MNRAS.433.3539G}
{Garrison-Kimmel} S., {Rocha} M., {Boylan-Kolchin} M., {Bullock} J.~S., {Lally}
  J., 2013, \mnras, 433, 3539

\bibitem[{{Goerdt} {et~al}\mbox{.}(2010){Goerdt}, {Moore}, {Read}, \&
  {Stadel}}]{2010ApJ...725.1707G}
{Goerdt} T., {Moore} B., {Read} J.~I., {Stadel} J., 2010, \apj, 725, 1707

\bibitem[{{Governato} {et~al}\mbox{.}(2014){Governato}, {Weisz}, {Pontzen},
  {Loebman}, {Reed}, {Brooks}, {Behroozi}, {Christensen}, {Madau}, {Mayer},
  {Shen}, {Walker}, {Quinn}, \& {Wadsley}}]{2014arXiv1407.0022G}
{Governato} F. {et~al.}, 2014, ArXiv e-prints

\bibitem[{{Gritschneder} \& {Lin}(2013)}]{2013ApJ...765...38G}
{Gritschneder} M., {Lin} D.~N.~C., 2013, \apj, 765, 38

\bibitem[{{Hague} \& {Wilkinson}(2014)}]{2014MNRAS.443.3712H}
{Hague} P.~R., {Wilkinson} M.~I., 2014, \mnras, 443, 3712

\bibitem[{{Hayashi} \& {Chiba}(2012)}]{2012ApJ...755..145H}
{Hayashi} K., {Chiba} M., 2012, \apj, 755, 145

\bibitem[{{Hinshaw} {et~al}\mbox{.}(2007){Hinshaw}, {Nolta}, {Bennett}, {Bean},
  {Dor{\'e}}, {Greason}, {Halpern}, {Hill}, {Jarosik}, {Kogut}, {Komatsu},
  {Limon}, {Odegard}, {Meyer}, {Page}, {Peiris}, {Spergel}, {Tucker}, {Verde},
  {Weiland}, {Wollack}, \& {Wright}}]{2007ApJS..170..288H}
{Hinshaw} G. {et~al.}, 2007, \apjs, 170, 288

\bibitem[{{Inoue} \& {Saitoh}(2011)}]{2011MNRAS.418.2527I}
{Inoue} S., {Saitoh} T.~R., 2011, \mnras, 418, 2527

\bibitem[{{Irwin} \& {Hatzidimitriou}(1995)}]{1995MNRAS.277.1354I}
{Irwin} M., {Hatzidimitriou} D., 1995, \mnras, 277, 1354

\bibitem[{{Ishiyama} {et~al}\mbox{.}(2013){Ishiyama}, {Rieder}, {Makino},
  {Portegies Zwart}, {Groen}, {Nitadori}, {de Laat}, {McMillan}, {Hiraki}, \&
  {Harfst}}]{2013ApJ...767..146I}
{Ishiyama} T. {et~al.}, 2013, \apj, 767, 146

\bibitem[{{Jardel} \& {Gebhardt}(2012)}]{2012ApJ...746...89J}
{Jardel} J.~R., {Gebhardt} K., 2012, \apj, 746, 89

\bibitem[{{Jing} \& {Suto}(2000)}]{2000ApJ...529L..69J}
{Jing} Y.~P., {Suto} Y., 2000, \apjl, 529, L69

\bibitem[{{Koch} {et~al}\mbox{.}(2007){Koch}, {Kleyna}, {Wilkinson}, {Grebel},
  {Gilmore}, {Evans}, {Wyse}, \& {Harbeck}}]{2007AJ....134..566K}
{Koch} A., {Kleyna} J.~T., {Wilkinson} M.~I., {Grebel} E.~K., {Gilmore} G.~F.,
  {Evans} N.~W., {Wyse} R.~F.~G., {Harbeck} D.~R., 2007, \aj, 134, 566

\bibitem[{{Komatsu} {et~al}\mbox{.}(2011){Komatsu}, {Smith}, {Dunkley},
  {Bennett}, {Gold}, {Hinshaw}, {Jarosik}, {Larson}, {Nolta}, {Page},
  {Spergel}, {Halpern}, {Hill}, {Kogut}, {Limon}, {Meyer}, {Odegard}, {Tucker},
  {Weiland}, {Wollack}, \& {Wright}}]{2011ApJS..192...18K}
{Komatsu} E. {et~al.}, 2011, \apjs, 192, 18

\bibitem[{{Lacey} \& {Cole}(1993)}]{1993MNRAS.262..627L}
{Lacey} C., {Cole} S., 1993, \mnras, 262, 627

\bibitem[{{Li} \& {White}(2008)}]{2008MNRAS.384.1459L}
{Li} Y.-S., {White} S.~D.~M., 2008, \mnras, 384, 1459

\bibitem[{{Ma} \& {Boylan-Kolchin}(2004)}]{2004PhRvL..93b1301M}
{Ma} C.-P., {Boylan-Kolchin} M., 2004, Physical Review Letters, 93, 021301

\bibitem[{{Madau}, {Shen} \& {Governato}(2014){Madau}, {Shen}, \&
  {Governato}}]{2014ApJ...789L..17M}
{Madau} P., {Shen} S., {Governato} F., 2014, \apjl, 789, L17

\bibitem[{{Majewski} {et~al}\mbox{.}(2003){Majewski}, {Skrutskie}, {Weinberg},
  \& {Ostheimer}}]{2003ApJ...599.1082M}
{Majewski} S.~R., {Skrutskie} M.~F., {Weinberg} M.~D., {Ostheimer} J.~C., 2003,
  \apj, 599, 1082

\bibitem[{{Mateo}(1998)}]{1998ARA&A..36..435M}
{Mateo} M.~L., 1998, \araa, 36, 435

\bibitem[{{Moore}(1994)}]{1994Natur.370..629M}
{Moore} B., 1994, \nat, 370, 629

\bibitem[{{Moore} {et~al}\mbox{.}(1999){Moore}, {Quinn}, {Governato}, {Stadel},
  \& {Lake}}]{1999MNRAS.310.1147M}
{Moore} B., {Quinn} T., {Governato} F., {Stadel} J., {Lake} G., 1999, \mnras,
  310, 1147

\bibitem[{{Mori} \& {Burkert}(2000)}]{2000ApJ...538..559M}
{Mori} M., {Burkert} A., 2000, \apj, 538, 559

\bibitem[{{Navarro}, {Eke} \& {Frenk}(1996){Navarro}, {Eke}, \&
  {Frenk}}]{1996MNRAS.283L..72N}
{Navarro} J.~F., {Eke} V.~R., {Frenk} C.~S., 1996, \mnras, 283, L72

\bibitem[{{Navarro}, {Frenk} \& {White}(1997){Navarro}, {Frenk}, \&
  {White}}]{1997ApJ...490..493N}
{Navarro} J.~F., {Frenk} C.~S., {White} S.~D.~M., 1997, \apj, 490, 493

\bibitem[{{Navarro} {et~al}\mbox{.}(2010){Navarro}, {Ludlow}, {Springel},
  {Wang}, {Vogelsberger}, {White}, {Jenkins}, {Frenk}, \&
  {Helmi}}]{2010MNRAS.402...21N}
{Navarro} J.~F. {et~al.}, 2010, \mnras, 402, 21

\bibitem[{{Niederste-Ostholt} {et~al}\mbox{.}(2010){Niederste-Ostholt},
  {Belokurov}, {Evans}, \& {Pe{\~n}arrubia}}]{2010ApJ...712..516N}
{Niederste-Ostholt} M., {Belokurov} V., {Evans} N.~W., {Pe{\~n}arrubia} J.,
  2010, \apj, 712, 516

\bibitem[{{Ogiya} \& {Mori}(2011)}]{2011ApJ...736L...2O}
{Ogiya} G., {Mori} M., 2011, \apjl, 736, L2

\bibitem[{{Ogiya} \& {Mori}(2014)}]{2014ApJ...793...46O}
{Ogiya} G., {Mori} M., 2014, \apj, 793, 46

\bibitem[{{Ogiya} {et~al}\mbox{.}(2014){Ogiya}, {Mori}, {Ishiyama}, \&
  {Burkert}}]{2014MNRAS.440L..71O}
{Ogiya} G., {Mori} M., {Ishiyama} T., {Burkert} A., 2014, \mnras, 440, L71

\bibitem[{{Oh} {et~al}\mbox{.}(2011){Oh}, {de Blok}, {Brinks}, {Walter}, \&
  {Kennicutt}}]{2011AJ....141..193O}
{Oh} S.-H., {de Blok} W.~J.~G., {Brinks} E., {Walter} F., {Kennicutt}, Jr.
  R.~C., 2011, \aj, 141, 193

\bibitem[{{Papastergis} {et~al}\mbox{.}(2014){Papastergis}, {Giovanelli},
  {Haynes}, \& {Shankar}}]{2014arXiv1407.4665P}
{Papastergis} E., {Giovanelli} R., {Haynes} M.~P., {Shankar} F., 2014, ArXiv
  e-prints

\bibitem[{{Pe{\~n}arrubia} {et~al}\mbox{.}(2012){Pe{\~n}arrubia}, {Pontzen},
  {Walker}, \& {Koposov}}]{2012ApJ...759L..42P}
{Pe{\~n}arrubia} J., {Pontzen} A., {Walker} M.~G., {Koposov} S.~E., 2012,
  \apjl, 759, L42

\bibitem[{{Pontzen} \& {Governato}(2012)}]{2012MNRAS.421.3464P}
{Pontzen} A., {Governato} F., 2012, \mnras, 421, 3464

\bibitem[{{Prada} {et~al}\mbox{.}(2012){Prada}, {Klypin}, {Cuesta},
  {Betancort-Rijo}, \& {Primack}}]{2012MNRAS.423.3018P}
{Prada} F., {Klypin} A.~A., {Cuesta} A.~J., {Betancort-Rijo} J.~E., {Primack}
  J., 2012, \mnras, 423, 3018

\bibitem[{{Press} \& {Schechter}(1974)}]{1974ApJ...187..425P}
{Press} W.~H., {Schechter} P., 1974, \apj, 187, 425

\bibitem[{{Richardson} \& {Fairbairn}(2013)}]{2013arXiv1305.0670R}
{Richardson} T., {Fairbairn} M., 2013, ArXiv e-prints

\bibitem[{{Salpeter}(1955)}]{1955ApJ...121..161S}
{Salpeter} E.~E., 1955, \apj, 121, 161

\bibitem[{{Salucci} \& {Burkert}(2000)}]{2000ApJ...537L...9S}
{Salucci} P., {Burkert} A., 2000, \apjl, 537, L9

\bibitem[{{Salvadori}, {Ferrara} \& {Schneider}(2008){Salvadori}, {Ferrara}, \&
  {Schneider}}]{2008MNRAS.386..348S}
{Salvadori} S., {Ferrara} A., {Schneider} R., 2008, \mnras, 386, 348

\bibitem[{{Sheth} \& {Tormen}(1999)}]{1999MNRAS.308..119S}
{Sheth} R.~K., {Tormen} G., 1999, \mnras, 308, 119

\bibitem[{{Spekkens}, {Giovanelli} \& {Haynes}(2005){Spekkens}, {Giovanelli},
  \& {Haynes}}]{2005AJ....129.2119S}
{Spekkens} K., {Giovanelli} R., {Haynes} M.~P., 2005, \aj, 129, 2119

\bibitem[{{Strigari} {et~al}\mbox{.}(2008){Strigari}, {Bullock}, {Kaplinghat},
  {Simon}, {Geha}, {Willman}, \& {Walker}}]{2008Natur.454.1096S}
{Strigari} L.~E., {Bullock} J.~S., {Kaplinghat} M., {Simon} J.~D., {Geha} M.,
  {Willman} B., {Walker} M.~G., 2008, \nat, 454, 1096

\bibitem[{{Strigari}, {Frenk} \& {White}(2014){Strigari}, {Frenk}, \&
  {White}}]{2014arXiv1406.6079S}
{Strigari} L.~E., {Frenk} C.~S., {White} S.~D.~M., 2014, ArXiv e-prints

\bibitem[{{Swaters} {et~al}\mbox{.}(2003){Swaters}, {Madore}, {van den Bosch},
  \& {Balcells}}]{2003ApJ...583..732S}
{Swaters} R.~A., {Madore} B.~F., {van den Bosch} F.~C., {Balcells} M., 2003,
  \apj, 583, 732

\bibitem[{{Takada}(2010)}]{2010AIPC.1279..120T}
{Takada} M., 2010, in American Institute of Physics Conference Series, Vol.
  1279, American Institute of Physics Conference Series, {Kawai} N., {Nagataki}
  S., eds., pp. 120--127

\bibitem[{{Tegmark} {et~al}\mbox{.}(2004){Tegmark}, {Blanton}, {Strauss},
  {Hoyle}, {Schlegel}, {Scoccimarro}, {Vogeley}, {Weinberg}, {Zehavi},
  {Berlind}, {Budavari}, {Connolly}, {Eisenstein}, {Finkbeiner}, {Frieman},
  {Gunn}, {Hamilton}, {Hui}, {Jain}, {Johnston}, {Kent}, {Lin}, {Nakajima},
  {Nichol}, {Ostriker}, {Pope}, {Scranton}, {Seljak}, {Sheth}, {Stebbins},
  {Szalay}, {Szapudi}, {Verde}, {Xu}, {Annis}, {Bahcall}, {Brinkmann},
  {Burles}, {Castander}, {Csabai}, {Loveday}, {Doi}, {Fukugita}, {Gott},
  {Hennessy}, {Hogg}, {Ivezi{\'c}}, {Knapp}, {Lamb}, {Lee}, {Lupton}, {McKay},
  {Kunszt}, {Munn}, {O'Connell}, {Peoples}, {Pier}, {Richmond}, {Rockosi},
  {Schneider}, {Stoughton}, {Tucker}, {Vanden Berk}, {Yanny}, {York}, \& {SDSS
  Collaboration}}]{2004ApJ...606..702T}
{Tegmark} M. {et~al.}, 2004, \apj, 606, 702

\bibitem[{{Teyssier} {et~al}\mbox{.}(2013){Teyssier}, {Pontzen}, {Dubois}, \&
  {Read}}]{2013MNRAS.429.3068T}
{Teyssier} R., {Pontzen} A., {Dubois} Y., {Read} J.~I., 2013, \mnras, 429, 3068

\bibitem[{{Tonini}, {Lapi} \& {Salucci}(2006){Tonini}, {Lapi}, \&
  {Salucci}}]{2006ApJ...649..591T}
{Tonini} C., {Lapi} A., {Salucci} P., 2006, \apj, 649, 591

\bibitem[{{Tsuchiya}, {Mori} \& {Nitta}(2013){Tsuchiya}, {Mori}, \&
  {Nitta}}]{2013MNRAS.432.2837T}
{Tsuchiya} M., {Mori} M., {Nitta} S.-y., 2013, \mnras, 432, 2837

\bibitem[{{van der Marel}(2006)}]{2006lgal.symp...47V}
{van der Marel} R.~P., 2006, in The Local Group as an Astrophysical Laboratory,
  {Livio} M., {Brown} T.~M., eds., pp. 47--71

\bibitem[{{van Eymeren} {et~al}\mbox{.}(2009){van Eymeren}, {Trachternach},
  {Koribalski}, \& {Dettmar}}]{2009A&A...505....1V}
{van Eymeren} J., {Trachternach} C., {Koribalski} B.~S., {Dettmar} R.-J., 2009,
  \aap, 505, 1

\bibitem[{{Walker} {et~al}\mbox{.}(2009){Walker}, {Mateo}, {Olszewski},
  {Pe{\~n}arrubia}, {Wyn Evans}, \& {Gilmore}}]{2009ApJ...704.1274W}
{Walker} M.~G., {Mateo} M., {Olszewski} E.~W., {Pe{\~n}arrubia} J., {Wyn Evans}
  N., {Gilmore} G., 2009, \apj, 704, 1274

\bibitem[{{Walker} \& {Pe{\~n}arrubia}(2011)}]{2011ApJ...742...20W}
{Walker} M.~G., {Pe{\~n}arrubia} J., 2011, \apj, 742, 20

\bibitem[{{Wang} {et~al}\mbox{.}(2012){Wang}, {Frenk}, {Navarro}, {Gao}, \&
  {Sawala}}]{2012MNRAS.424.2715W}
{Wang} J., {Frenk} C.~S., {Navarro} J.~F., {Gao} L., {Sawala} T., 2012, \mnras,
  424, 2715

\bibitem[{{Wolf} {et~al}\mbox{.}(2010){Wolf}, {Martinez}, {Bullock},
  {Kaplinghat}, {Geha}, {Mu{\~n}oz}, {Simon}, \& {Avedo}}]{2010MNRAS.406.1220W}
{Wolf} J., {Martinez} G.~D., {Bullock} J.~S., {Kaplinghat} M., {Geha} M.,
  {Mu{\~n}oz} R.~R., {Simon} J.~D., {Avedo} F.~F., 2010, \mnras, 406, 1220

\end{thebibliography}

\label{lastpage}
\end{document}